\documentclass[preprintnumbers,amsmath,amssymb,nofootinbib]{revtex4}
\usepackage{epsfig}
\usepackage{bm}

\newcommand{\s}{\sigma}
\newcommand{\la}{\lambda}

\begin{document}

\title{Classical evolution of fractal measures generated by a scalar field on the lattice}

\author{N.~G.~Antoniou}
\author{F.~K.~Diakonos}
\author{E.~N.~Saridakis}
\email{msaridak@phys.uoa.gr}
\author{G.~A.~Tsolias}
\affiliation{Department of Physics, University of Athens, GR-15771
Athens, Greece}

\date{\today}

\begin{abstract}

We investigate the classical evolution of a $\phi^4$ scalar field
theory, using in the initial state random field configurations
possessing a fractal measure expressed by a non-integer mass
dimension. These configurations resemble the equilibrium state of
a critical scalar condensate. The measures of the initial fractal
behavior vary in time following the mean field motion. We show
that the remnants of the original fractal geometry survive and
leave an imprint in the system time averaged observables, even for
large times compared to the approximate oscillation period of the
mean field, determined by the model parameters. This behavior
becomes more transparent in the evolution of a deterministic
Cantor-like scalar field configuration. We extend our study to the
case of two interacting scalar fields, and we find qualitatively
similar results. Therefore, our analysis indicates that the
geometrical properties of a critical system initially at
equilibrium could sustain for several periods of the field
oscillations in the phase of non-equilibrium evolution.

\end{abstract}

\maketitle

\section{Introduction}

The classical dynamics of scalar field theory has been extensively
studied in the literature. In most investigations a lattice
discretization of the scalar field is used, reducing the problem
to the study of the dynamics of a system of non-linear coupled
oscillators \cite{Toda89}. The central question in this case is
the evolution of the system towards a thermalized stationary
state. In the early days Fermi, Pasta and Ulam \cite{Fermi55} have
obtained deviations, even for large times, from the naively
expected equipartition of the energy among the different
oscillators. Through the efforts to explain these results, it
became clear that, for appropriate initial conditions, a variety
of periodic solutions (breathers, solitary waves) \cite{Ford92},
defined on the non-linear lattice, exists. Therefore, the choice
of the ensemble of the initial configurations strongly influences
the long time behavior of the system dynamics. Recent works
\cite{Parisi97,Wetter99} show that for a random ensemble of
initial configurations a sufficiently large system relaxes to the
usual equilibrium distribution, but the corresponding relaxation
time strongly depends on the parameters of the theory.

In the present work we reconsider the classical dynamics of a
scalar field in $1+1$ dimensions adopting a different point of
view: As initial conditions we use an ensemble of scalar field
configurations possessing a non-conventional profile, inspired by
the order parameter fluctuations of a critical system at thermal
equilibrium. These configurations generate a fractal measure on
the lattice characterized  by a corresponding fractal mass
dimension. We do not consider here the dynamical process
responsible for the formation of such a critical state, but we
concentrate on its evolution once it has been formed. Our aim is
to investigate the deformation of the initial fractal measure as
the system evolves according to the classical equations of motion.
In particular, we are interested in determining the time scale for
which signatures of the initial fractality survive, and leave
their imprint in appropriate observables. We find that the initial
geometry is successively deformed and restored again, with a
frequency determined by the field oscillations. This behavior
seems to be generic, since it is observed for both random as well
as deterministic (Cantor-like) fractal measure. Moreover, we study
in detail the influence of an additional thermalized non-critical
scalar field, described initially by configurations with
conventional geometry, coupled to the system, as well as the
dependence of the corresponding characteristic time scales on the
parameters of the theory. Finally, we discuss the applicability of
our model to the out-of-equilibrium evolution of an isoscalar
condensate formed near the Quantum Chromodynamics (QCD) critical
point during a heavy-ion collision experiment.

The paper is organized as follows: Section II is divided in two
subsections. In subsection IIA we present our model considering a
single self-interacting scalar field. In IIB we describe the
generation of the ensemble of initial configurations for a scalar
field corresponding to a random fractal measure on the lattice
with a given fractal mass dimension. Section III contains the
numerical results of the single field case and section IV the
corresponding results for the evolution of field configurations
initially generating a deterministic fractal measure with
Cantor-like structure. In section V we analyze the dynamics of two
coupled scalar fields. Finally, in section VI we summarize our
results and discuss their relevance to the phenomenology of
out-of-equilibrium critical systems.

\section{The Model}

\subsection{Equations of motion}

Our model consists of a classical scalar field $\s$ obeying the
usual $\phi^4$-dynamics described by the Lagrangian density:
\begin{equation}
\mathcal{L}=\frac{1}{2}\partial_\mu\s\partial^\mu\s-V(\s)
\label{lagrs}
\end{equation}
 with the potential
\begin{equation}
V(\s)=\frac{\la}{4}(\s^2-1)^2-A\s,
 \label{pots}
\end{equation}
where  $\la$ and $A$ are the coupling parameters of our model. All
the quantities ($\s$, $\la$, $A$, as well as the space-time
variables) appearing above are chosen dimensionless. Following the
$\s$-model we assume that the $Z_2$ symmetry ($\s\rightarrow-\s$)
is broken only through a linear term in the potential, setting the
coefficient of the cubic term to zero. Furthermore, we have
absorbed one more parameter by rescaling the field as well as the
space-time units. Thus, only two parameters remain in the
potential term. We consider the dynamics of the scalar field in
$1+1$ dimensions. The corresponding equation of motion is
\begin{eqnarray}
\ddot{\s}-\s''+\la\s^3-\la\s-A=0\label{eoms}
\end{eqnarray}
where dot represents time derivative and prime the spatial one. To
proceed numerically we have to discretize eq.(\ref{eoms}) on a
lattice. This reduces the system to a chain of non-linear coupled
oscillators. We use the following leap-frog discretization scheme:
\begin{equation}
\s^{n+2}_i=2\s^{n+1}_i-\s^{n}_i+\frac{dt^2}{dx^2}
\left(\s^{n+1}_{i+1}+\s^{n+1}_{i-1}-2\s^{n+1}_{i}\right)-
dt^2\left[\lambda(\s^{n+1}_{i})^3-\lambda\s^{n+1}_{i}-A\right],
\label{leap-frog}
\end{equation}
where $dx$ is the lattice spacing, $dt$ is the time step, the
upper indices correspond to time steps and the lower indices to
lattice sites. As usual we
 perform an initial fourth order
Runge-Kutta step to make our algorithm self-starting. We are
interested in studying the evolution of the above system
determined by eq.(\ref{eoms}), using an ensemble of initial field
configurations possessing a non-conventional profile,
characterized by a fractal mass dimension. The motivation of this
choice and the details of constructing such an ensemble, defined
on a 1-dimensional lattice, are given in the next subsection.

\subsection{Generation of initial ensemble of $\sigma$-configurations corresponding to a random fractal measure}

The absolute value of the $\s$-field introduced in the previous
subsection is interpreted as local density, and the corresponding
fractal behavior is described by a fractal measure demonstrated in
the dependence of the mean "mass" on the distance $R$ around a
point $\vec{x}_0$ defined by:
\begin{equation}
m(\vec{x}_0,R)=\langle\int_{R} |\sigma(\vec{x}-\vec{x}_0)|\,
d^Dx\rangle, \label{massdimen}
\end{equation}
obeying the power law
\begin{equation}
m(\vec{x}_0,R)\sim R^{D_f} \label{masspower}
\end{equation}
for every $\vec{x}_0$. $D_f$ is the fractal mass dimension of the
system \cite{Mandel83,Vicsek,Falconer} and the mean value is taken
with respect to the ensemble of the initial $\s$-configurations.
The production of a $\s$-ensemble possessing the fractal measure
described in eqs.~(\ref{massdimen},\ref{masspower}), has been
accomplished in \cite{Antoniou98}. It is based on the observation
that a scale invariant free energy of the form:
\begin{equation}
\Gamma[\sigma]=\int_V d^Dx \{ \frac{1}{2} (\nabla \sigma)^2 + g
\sigma^{\delta+1}  \}, \label{effact}
\end{equation}
 when introduced as a weight in the partition function:
\begin{equation}
Z=\int {\cal{\delta}}[\sigma] e^{-\Gamma[\sigma]},\nonumber
\end{equation}
of the scalar field $\s$, generates piecewise constant
configurations leading to an ensemble possessing fractal mass
dimension (according to the definition above):
\begin{equation}
D_f=\frac{D\delta}{\delta+1} \label{fracdim}.
 \end{equation}
In the following we will use $\delta=5$ and the dimension $D=1$,
therefore the corresponding fractal mass dimension is $D_f=5/6$.
It must be noted that the free energy (\ref{effact}) for $D=3$,
$\delta=5$ and $g=2$ describes the effective action of the $3D$
Ising model at its critical point \cite{Tsypin}.

In practice, to produce critical configurations on a lattice we
use the following algorithm \cite{algorithm}: We perform a random
partitioning of the lattice in elementary clusters of different
size $\xi$. Thus, each cluster consists of several lattice points.
Within each cluster the value of $\s$ is assumed to be constant.
To obtain the values of $\s$ in the different clusters we use a
uniform random distribution. Each $\s$-field configuration is
weighted by a factor $e^{-\Gamma[\sigma]}$, where $\Gamma[\sigma]$
is calculated for the given configuration, using the coupling
value $g=2$. We use a Metropolis algorithm \cite{Metropolis55} to
perform a random walk in the configuration space of the
$\s$-field. We perform firstly $\approx 25000$ initial algorithmic
steps in order to achieve equilibrium. The $\sigma$ ensemble is
then formed by recording a large number ($\sim10^4$) of
statistically independent $\s$-configurations.

\begin{figure}[h]
\begin{center}
\mbox{\epsfig{figure=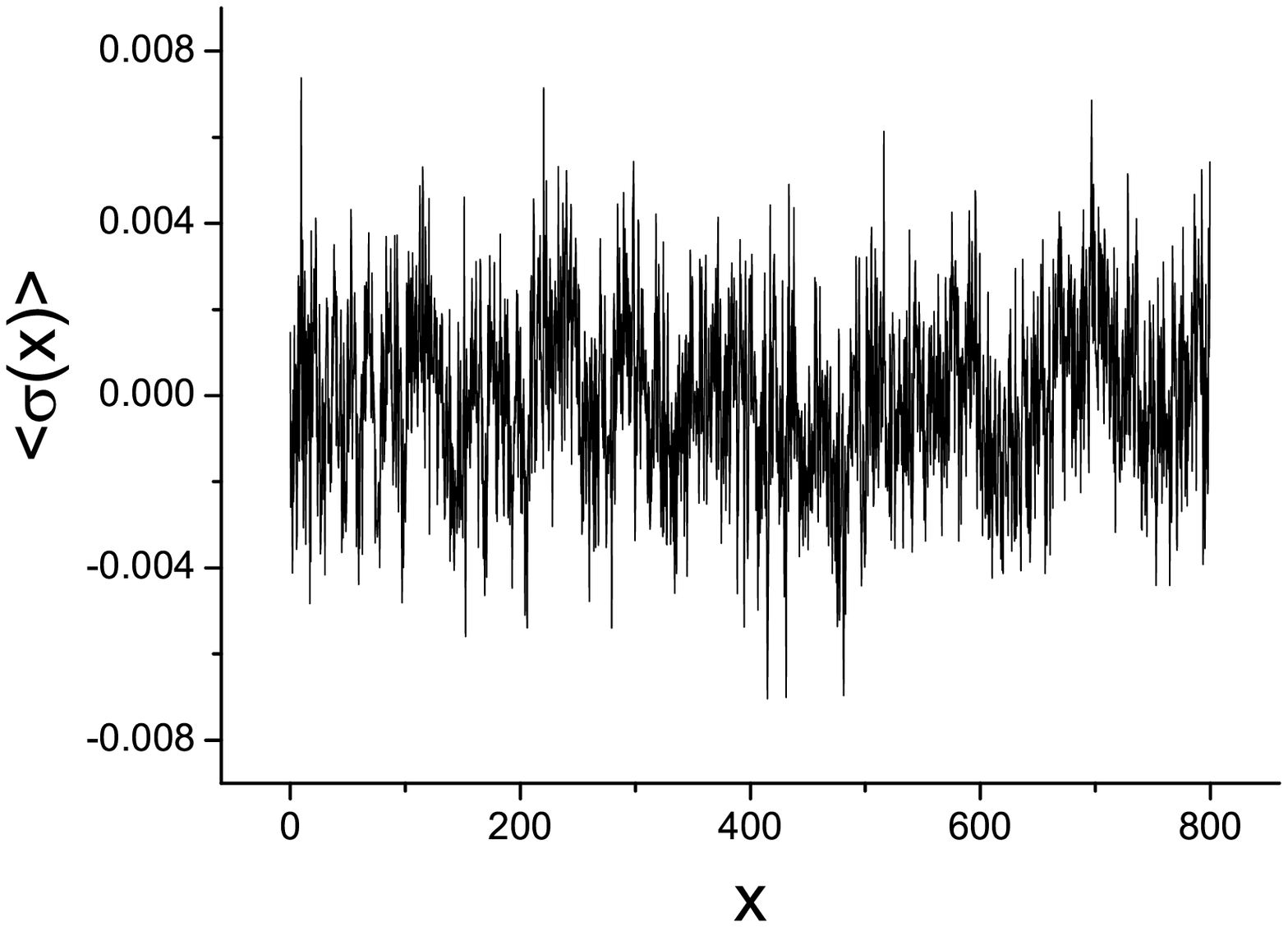,width=9cm,angle=0}}
\caption{\it The $\s$-field on the lattice averaged over the
ensemble of the initial configurations.} \label{sigma0}
 \end{center}
 \end{figure}

With this procedure we acquire an ensemble of field configurations
generating a random fractal measure on the lattice as a
statistical property after ensemble averaging. This property is
not reflected in the geometry of either one configuration neither
in their average $\langle\sigma(x)\rangle$, which is depicted in
fig.~\ref{sigma0}, but is produced only through the entire
ensemble. Each one of the configurations as well as their average,
have a continuous power spectrum ($P(f)\sim f^{-1/2}$) resembling
a colored noise profile, and the mean value of the field (spatial
average) is almost zero.

The produced ensemble possesses the property (\ref{masspower}),
where now the fractal mass dimension is determined by the
power-law behavior of $m(x_0,\xi)=\langle\int^{x_0+\xi}_{x_0}
|\sigma(x)| dx\rangle$ around a random $x_0$, averaged inside
clusters of size $\xi$ \footnote{Note that this fractal mass
dimension must not be confused with the fractal dimension of the
corresponding curve, which in this case is greater than 1.}. The
$\langle\int^{x_0+\xi}_{x_0} |\sigma(x)| dx\rangle$ versus $\xi$
figure is drawn as follows: For a given $x_0$ of a specific
configuration we find the size $\xi$ of the cluster in which it
belongs and we calculate the integral $\int^{x_0+\xi}_{x_0}
|\sigma(x)| dx$, thus acquiring one point in the
$\langle\int^{x_0+\xi}_{x_0} |\sigma(x)| dx\rangle$ vs $\xi$
figure. For the same $x_0$ we repeat this procedure until we cover
the whole ensemble, and the aforementioned figure is formed.
Averaging in $x_0$ obviously does not alter the results, since
$m(x_0,\xi)\approx m(x_0+l,\xi)$, with $l$ spanning the entire
lattice. In fig \ref{spower1}a we observe that in the log-log plot
of $\langle\int^{x_0+\xi}_{x_0} |\sigma(x)| dx\rangle$ vs $\xi$,
the slope $\psi$, i.e the fractal mass dimension $D_f$ according
to (\ref{masspower}),
 is equal to $5/6$, which is the theoretical value
calculated from (\ref{fracdim}), within an error of less than
0.3\%.
\begin{figure}[h]
\begin{center}
\mbox{\epsfig{figure=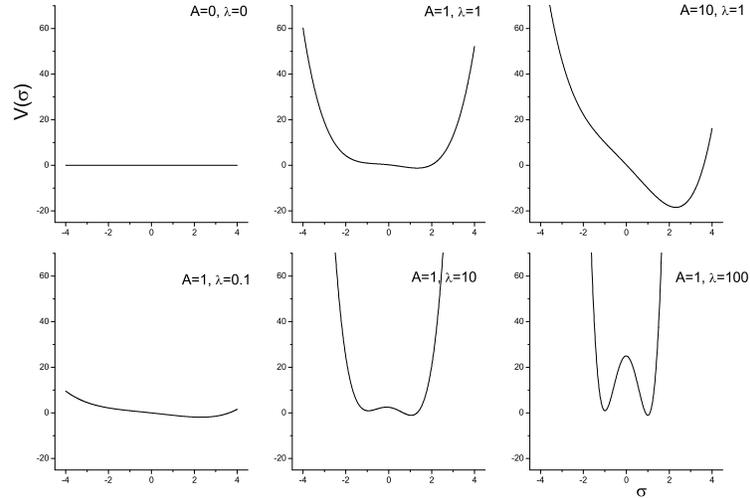,width=11cm,angle=0}}
\caption{\it Potential $V(\s)$ given by (\ref{pots}), for
different values of $A$ and $\la$.} \label{potetial.plot}
 \end{center}
 \end{figure}
 \begin{figure}[h]
\begin{center}
\mbox{\epsfig{figure=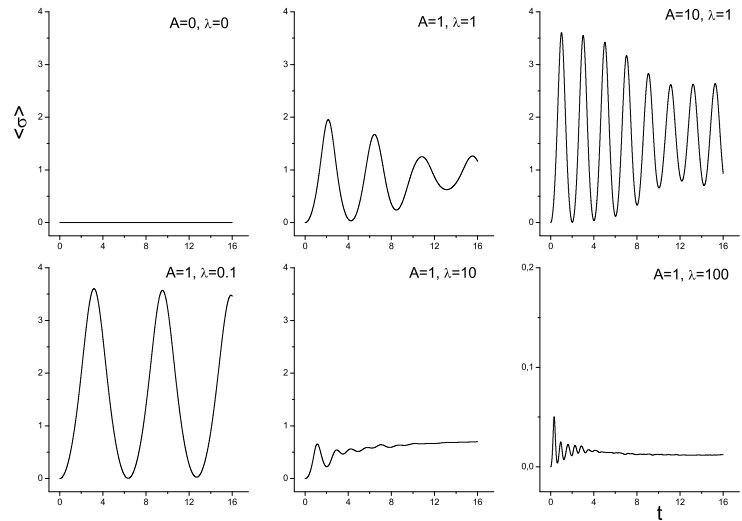,width=11cm,angle=0}}
\caption{\it Time evolution of the ensemble average of the spatial
mean $\langle\s\rangle$ for different values of $A$ and $\la$ in
dimensionless units. Mind the different vertical axis scale in the
last graph.} \label{sketo.s.final1}
 \end{center}
 \end{figure}

\section{Numerical Results. Single field}

We study the evolution of the system determined by equation
(\ref{eoms}) which we solve in 1-D 2000-site lattice, using as
initial conditions an ensemble of $10^4$ independent
$\s$-configurations on the lattice generated as described above,
i.e possessing fractal characteristics. The initial time
derivatives of the field, that is the kinetic energy, are assumed
to be zero, since this is a strong requirement of the initial
equilibrium. The ensemble population is by far satisfactory since
the results are independent of it as long as it is larger than
$6\times10^3$ (numerically tested), and furthermore they are
independent from the number of the lattice sites provided that it
is greater than $\sim10^3$. We investigate the evolution of
$m(x_0,\xi)(\ =\langle\int^{x_0+\xi}_{x_0} |\sigma(x)| dx\rangle)$
which initially is a power law $\sim\xi^{\psi(0)}$, where
$\psi(0)=D_f=5/6$ (fractal mass dimension).

\begin{figure}[h]
\begin{center}
\mbox{\epsfig{figure=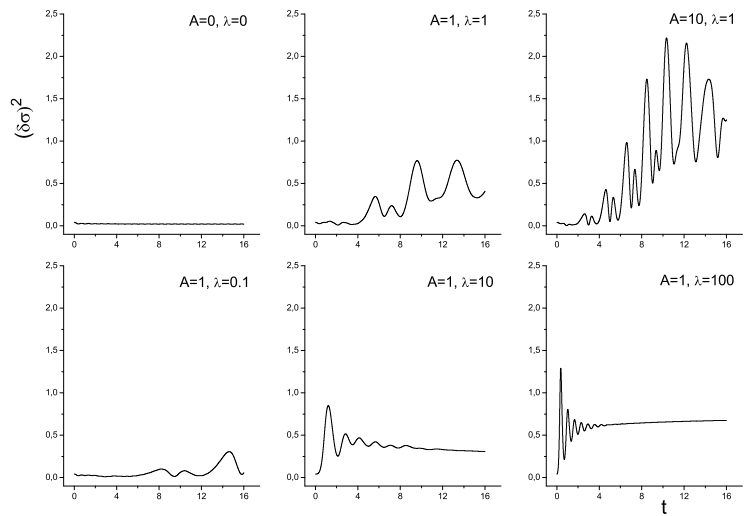,width=11cm,angle=0}}
\caption{\it The standard deviation time evolution of the ensemble
average of the $\s$-field for different values of $A$ and
 $\la$ in dimensionless units.}
\label{sketo.s.final2}
 \end{center}
 \end{figure}

In order to understand the dynamics of the $\s$-field within this
ensemble we firstly consider the evolution of the field's mean
value $\langle\s\rangle$, as well as its standard deviation
$(\delta\s)^2=\langle\s^2\rangle-\langle\s\rangle^2$. The averages
are taken over all statistically independent configuration. We
classify the dynamics in six different cases, produced through a
suitable choice of the parameters $A$ and $\lambda$, according to
the corresponding form of the potential plotted in fig
\ref{potetial.plot}. In fig.~\ref{sketo.s.final1} we depict the
 evolution of $\langle\s\rangle$ for these cases. For zero $A$ and
 $\la$, the potential term vanishes and the mean value of the field remains constant and equal to a very small value
 determined by the initial conditions.
For non-zero $A$ the field oscillates around the potential minimum
and the oscillation amplitude as long as the frequency increase
with $A$ for fixed $\la$. This is due to the fact that the minimum
value of the potential decreases and at the same time the value of
$\s$, for which the potential minimum occurs, increases (see
figs.~\ref{potetial.plot} and \ref{sketo.s.final1}).
 Note furthermore that due to the quadric anharmonic
term in the potential the time mean value of the oscillations is
slightly smaller than the potential minimum. On the other hand,
for small $\la$, $\langle\s\rangle$ oscillates almost
harmonically, and the amplitude together with the minimum decrease
with increasing $\la$, while the frequency increases. For $\la>1$
the oscillations damp relatively fast (the larger the $\la$ values
the faster the rate, as the potential becomes steeper), leading to
a stabilization of $\langle\s\rangle$.
 Finally, in fig.~\ref{sketo.s.final2} we show  the evolution of the ensemble average
of the standard deviation of the $\s$-field for the same
parameters as above. Note that the standard deviation can be quite
large even if $\langle\s\rangle$ remains small, such as in the
$A=1$, $\la=100$ case.

\begin{figure}[h]
\begin{center}
\mbox{\epsfig{figure=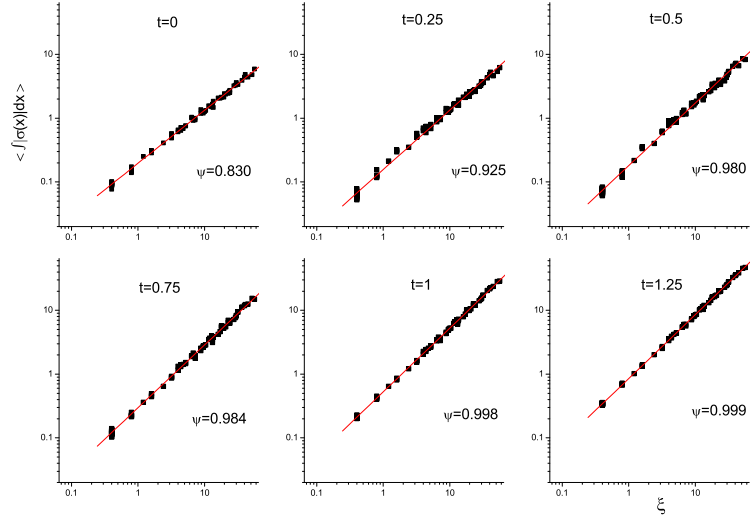,width=11cm,angle=0}}
\caption{\it $\langle\int^{x_0+\xi}_{x_0} |\sigma(x)| dx\rangle$
versus $\xi$, for successive times $t=0$, $t=0.25$, $t=0.5$,
$t=0.75$, $t=1$, $t=1.25$, for $A=1$ and $\la=1$ in dimensionless
units. We observe that although the initial fractal mass dimension
changes, the power law property remains valid as time evolves.}
\label{spower1}
 \end{center}
 \end{figure}

\begin{figure}[h]
\begin{center}
\mbox{\epsfig{figure=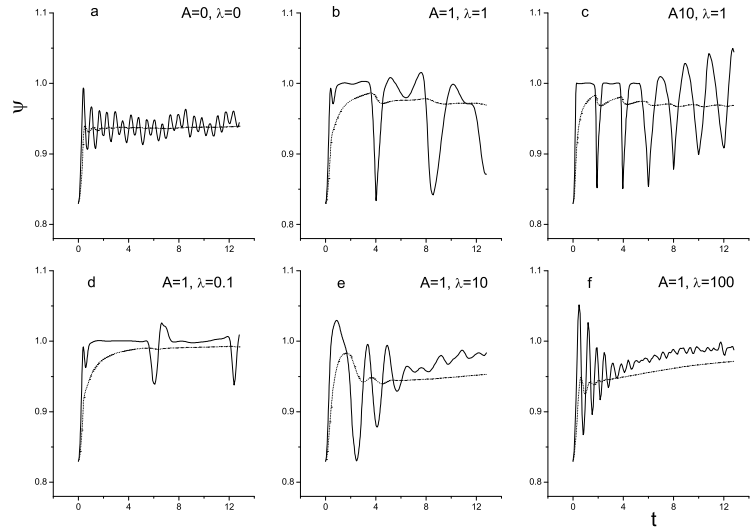,width=11cm,angle=0}}
\caption{\it Time evolution of the slope $\psi$ (solid line) and
its time average $\langle\psi\rangle_t$ (dotted line) for varied
$A$ and $\la$ in dimensionless units.} \label{spower.evol}
 \end{center}
 \end{figure}

As a next step we calculate the evolution of $m(x_0,\xi)$ defined
previously. As time passes the power-law form of $m(x_0,\xi)\
(\sim\xi^{\psi(t)})$ remains, but the corresponding exponent
$\psi(t)$ increases, approaching the value $\psi(t)\approx1$, when
the signs of the initial fractal geometry disappear and a
conventional pattern establishes, as is depicted in
fig.~\ref{spower1}.
 However, a more detailed analysis
for greater time intervals reveals a remarkable phenomenon.
 In the solid line plots of fig.~\ref{spower.evol} we show the evolution of $\psi$
  (each $\psi$ value coming from a linear fit).
It can be clearly observed that the characteristic exponent
$\psi(t)$ after reaching the value 1, fluctuates and for
particular times becomes almost equal to $\psi(0)=5/6$. Thus, the
initial fractal measure is restored repeatedly within the
considered ensemble. A simultaneous study of this graph and
fig.~\ref{sketo.s.final1} is illuminating, since we observe that
the reappearance of the fractal mass dimension, describing the
initial measure, occurs when the mean field value becomes almost
zero.

 If we interpret the $\s$-field as a
system of coupled anharmonic oscillators with initial field values
around zero and zero kinetic energy, we expect an almost
simultaneous pass from their turning points, due to
synchronization. The energy transfer between the different
oscillators takes place through the spatial derivative, as can be
seen in equation (\ref{eoms}). Therefore, if it is small compared
to the other terms (which is the case in general), the oscillators
do not mix significantly, and due to their initial zero kinetic
energy they roll to their common minimum, they oscillate around it
and then return to their starting position almost in phase and
with kinetic energy close to the initial one, i.e close to zero.
 So, every time the kinetic energy and the mean
field value become zero, the system reaches to a state similar to
the initial one, leading to the restoration of the initial fractal
measure. On the other hand, when the oscillators roll away from
zero, the geometrical characteristics of the ensemble are changed
by the dynamics, since they acquire large field values and large
kinetic energy (compared to the initial ones).

 In figs
\ref{spower.evol}b,c,d we can clearly observe this behavior.
Moreover, $\psi(t)$ in figs \ref{spower.evol}e and f is also
easily interpreted since in these cases, following closely the
evolution of the $\s$ mean value, it stabilizes after some
reciprocations. Finally, the behavior of $\psi$ for $A=0$, $\la=0$
depicted in fig.~\ref{spower.evol}a is also expected since in this
case (\ref{eoms}) reduces to the simple wave equation, $\s$ values
stay always around zero, i.e around their initial values, and
$\psi$ oscillates around a value between $5/6$ and $1$, namely the
signs of the initial fractal geometry are always visible. The
dotted line plots of fig.~\ref{spower.evol} depict the time
average of $\psi$ defined by
$\langle\psi\rangle_t=\frac{1}{t}\int_0^t\psi(t')dt'$. As it can
be observed, the influence of the initial fractal mass dimension
is still visible in this time integrated measure.

\begin{figure}[h]
\begin{center}
\mbox{\epsfig{figure=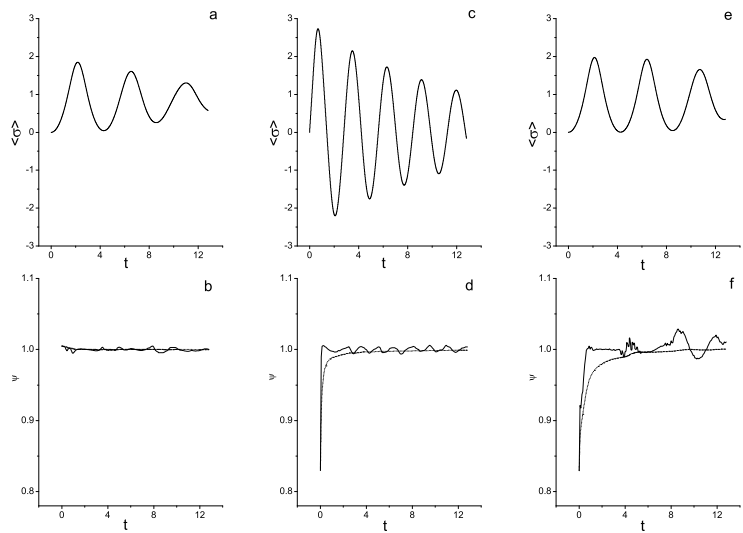
,width=11cm,angle=0}} \caption{\it Mean field, $\psi(t)$ (solid
line) and  $\langle\psi\rangle_t$ (dotted line) evolution for
various scenarios. a) and b) correspond to random non-fractal
initial conditions for $A=1$ and $\la=1$, c) and d) correspond to
 initial conditions with fractal mass dimension $\psi(0)=5/6$ and non-zero, randomly chosen, initial
kinetic energy, for $A=1$ and $\la=1$, and e) and f) correspond to
 initial conditions with fractal mass dimension $\psi(0)=5/6$, zero kinetic energy, but three
orders of magnitude larger spatial derivative term, for $A=1$ and
$\la=1$.} \label{DIAFORSTESTS}
 \end{center}
 \end{figure}

Lastly, in order to have a clearer apprehension of the
aforementioned phenomenon, we perform some tests. In fig.
\ref{DIAFORSTESTS}a and \ref{DIAFORSTESTS}b we present the
evolution of the system with random initial conditions, prepared
choosing the $\s$ value at each site from a uniform distribution,
for $A=1$ and $\la=1$. The mean field value oscillates around the
minimum as before, and the slope $\psi$ which initially is
obviously one, corresponding to a non-fractal system, remains
equal to 1 as expected, independently of the field motion.
Following the arguments referred above, one could say that every
time $\langle\s\rangle$ returns to zero the system enters in a
state similar to the initial one, and $\psi(t)$ remains always 1
since the initial state is characterized by $\psi(0)=1$. In figs
\ref{DIAFORSTESTS}c and \ref{DIAFORSTESTS}d we evolve our system
using initial conditions corresponding to a fractal measure with
mass dimension $\psi(0)=5/6$, but with random non-zero (actually
quite large) kinetic energy, for $A=1$, $\la=1$. In this case the
finite value of the initial kinetic energy (different for every
oscillator) forbids the return of the system to a state close to
the initial one, each time the mean value of the field passes
through zero, suppressing the approach to the initial fractal
state. (Note that due to the large initial kinetic energy the
system oscillates around both minima).

Additionally, in figs \ref{DIAFORSTESTS}e and \ref{DIAFORSTESTS}f
we perform the following scenario: We evolve the initial ensemble,
possessing fractal mass dimension $\psi(0)=5/6$, for $A=1$,
$\la=1$, introducing by hand  a three orders of magnitude larger
coefficient to the spatial derivative term in the equation of
motion. As expected, the enhanced diffusion in this case induces
very strong mixing of the oscillators, and as a consequence the
system never reacquires the initial fractal characteristics
\footnote{
 Note that the increased diffusion can also be achieved using the
 modified form of the potential:
 $V(\s)=\frac{\la}{4}(\s^2-v^2)^2-A\s$, taking small $A$ and
$\la$ and large $v$.}.

\section{Evolution of a scalar field configuration with deterministic fractal measure}

In order to acquire a better comprehension of the mechanism of the
aforementioned phenomenon, we investigate the evolution of one
Cantor-like scalar field configuration. This set up, although not
related to critical phenomena, is enlightening since in this case
a direct geometrical interpretation at the level of one
configuration is possible, instead of investigating the
statistical fractal properties of an ensemble.

\begin{figure}[h]
\begin{center}
\mbox{\epsfig{figure=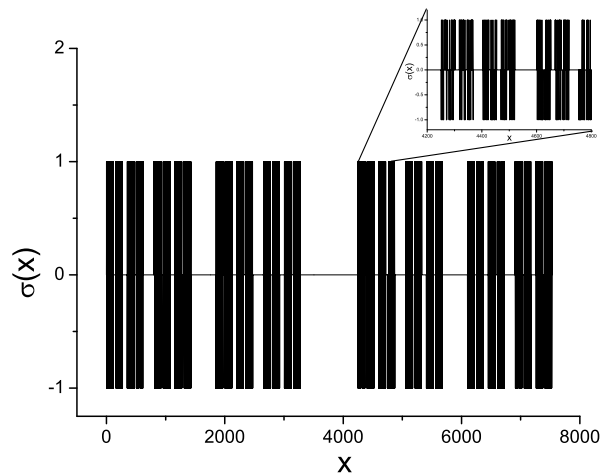 ,width=9cm,angle=0}}
\caption{\it The $\s$-field derived by a Cantor-like fractal, on
the equidistant lattice.} \label{Det.sav}
 \end{center}
 \end{figure}

Firstly, we construct a finite approximation to the 1D Cantor dust
of $2^{11}$ sites, with Hausdorf fractal dimension $D_f=5/6$
\cite{Mandel83,lala}. In order to transform this fractal set into
a field configuration defining a fractal measure on an equidistant
lattice we find the minimum two-point distance of the set and
using it as the lattice spacing we determine the size of the
lattice by dividing the maximum two point distance in the set by
the minimum one. In the sites of this new equidistant lattice that
are closer to the locations of the points of the initial Cantor
set, we give the field value 1, while in all the others we give
the field value 0. Thus, we turn out with a field configuration in
an equidistant lattice, which by construction has the property
$\langle\int^{x_{0,i}+\zeta}_{x_{0,i}} |\sigma(x)|
dx\rangle\propto\zeta^{D_f}$ in a good precision, where the
reference sites $x_{0,i}$ ($i=1,...,2^{11}$) are obviously only
those with field values 1. The averaging now is taken only on the
different $x_{0,i}$ since we have only one configuration. The
produced $\s$-field configuration is depicted in
fig.~\ref{Det.sav} where the fractal property is clear. Note that
starting from the initial $2^{11}$ Cantor lattice, we transited in
a much larger ($\approx2\times10^4$ sites) equidistant one.

\begin{figure}[!]
\begin{center}
\mbox{\epsfig{figure=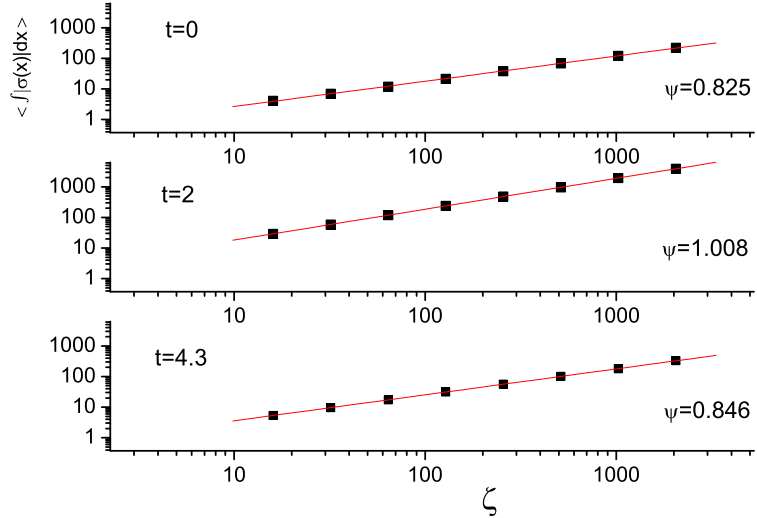 ,width=11cm,angle=0}}
\caption{\it $\langle\int^{x_{0,i}+\zeta}_{x_{0,i}} |\sigma(x)|
dx\rangle$ versus $\zeta$,  for successive times $t=0$ (initial
moment), $t=2$ (corresponding to the first complete destruction of
the initial power law), and $t=4.3$ (corresponding to the first
restoration of the initial fractal measure), for $A=1$ and
$\la=1$, in dimensionless units.} \label{Det.3power}
 \end{center}
 \end{figure}

Now, we evolve this $\s$-configuration according to equation of
motion (\ref{eoms}) taking zero initial kinetic energy, and we
focus on the evolution of $\langle\int^{x_{0,i}+\zeta}_{x_{0,i}}
|\sigma(x)| dx\rangle$, which initially has the characteristic
fractal mass dimension $\psi(0)=D_f=5/6$ as can be seen in the
upper graph of fig.~\ref{Det.3power}.
 \begin{figure}[h]
\begin{center}
\mbox{\epsfig{figure=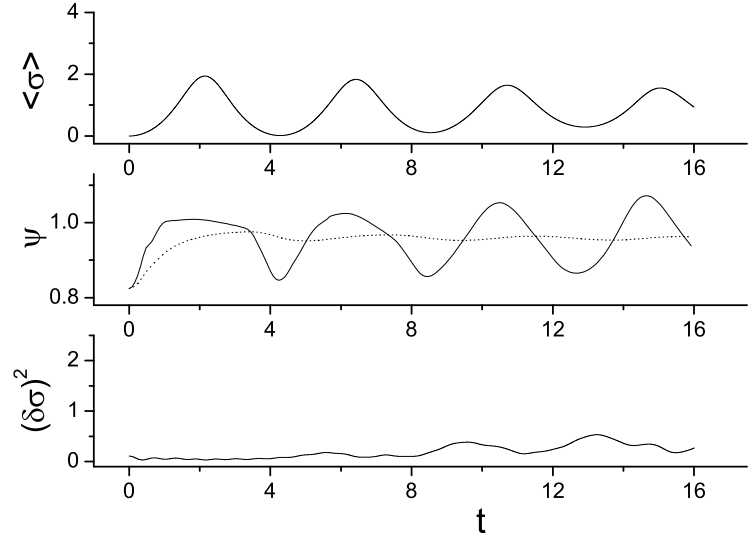
,width=11cm,angle=0}} \caption{\it Mean field, standard deviation
and  $\psi$ (solid line) and $\langle\psi\rangle_t$ (dotted line)
evolution for the deterministic fractal case, for $A=1$ and
$\la=1$, in dimensionless units.} \label{Det.fields}
 \end{center}
 \end{figure}
In fig.~\ref{Det.fields} we demonstrate the evolution of the mean
value of the field, its standard deviation, and the slope $\psi$,
for potential parameters $A=1$ and $\lambda=1$. We observe the
same behavior with the case of the previous section, that is the
change of the initial fractal mass dimension, and its return to
this value each time $\langle\s\rangle$ approaches zero. The
explanation is the same as before without any significant new
ideas. The only difference is that in the present case only a
small subset of the entire lattice is initially described by a
fractal measure with mass dimension $5/6$. The remaining lattice
sites are occupied by zero field values, which initially have zero
contribution to $\langle\int^{x_{0,i}+\zeta}_{x_{0,i}} |\sigma(x)|
dx\rangle$. However, as the system evolves, these initially zero
oscillators roll towards the potential minimum, producing a
non-fractal background, i.e a linear contribution to
$\langle\int^{x_{0,i}+\zeta}_{x_{0,i}} |\sigma(x)| dx\rangle$, and
thus supplanting and suppressing the total power law behavior. On
the other hand, when the mean field value returns to zero, the
system reaches a state similar to the initial one, (zero mean
field value and zero kinetic energy), the initially zero
oscillators approach again the value $\s=0$ almost in phase and
reproduce a configuration with fractal mass dimension $D_f\approx
5/6$. In fig.~\ref{Det.3power} we  depict
$\langle\int^{x_{0,i}+\zeta}_{x_{0,i}} |\sigma(x)|dx\rangle$
versus $\zeta$ for three times. The first corresponds to the
initial moment, the second to the first complete destruction of
the fractal geometry, and the third to the first approximate
re-establishment of the initial fractal measure.

\section{Two coupled fields}

It is interesting to extend the single field model of section II
in the case of two coupled fields, one of them having a fractal
profile and the other a conventional one. In this case the
Lagrangian density (\ref{lagrs}) is extended to
\begin{equation}
\mathcal{L}=\frac{1}{2}(\partial_\mu\s\partial^\mu\s
+\partial_\mu\pi\partial^\mu\pi)-V(\s,\pi), \label{lagrsp}
\end{equation}
 with the potential
\begin{equation}
V(\s,\pi)=\frac{\la}{4}(\s^2+\pi^2-1)^2-A\s, \label{potsp}
\end{equation}
avoiding to introduce additional parameters. The fields are
defined in $1+1$ dimensions. The equations of motion derived from
(\ref{lagrsp}) are
\begin{eqnarray}
\ddot{\s}-\s''+\la\s^3+\la(\pi^2-1)\s-A=0\nonumber\\
\ddot{\pi}-\pi''+\la\pi^3+\la(\s^2-1)\pi=0. \label{eomsp}
\end{eqnarray}

We are interested in studying the evolution of our system
determined by the above equations, taking for the $\s$-field
initial conditions possessing fractal behavior as described in
section II, and for the $\pi$-field conventional initial
conditions, corresponding to an ideal gas at temperature $T_0$. In
the next subsection we describe the generation of the ensemble of
1-d $\pi$-configurations on the lattice, using the algorithm
introduced in \cite{Cooper}.

\subsection{Generation of thermal $\pi$-configurations}

 The unperturbed Hamiltonian for the classical scalar field
theory in 1-d is
\begin{equation}
H=\frac{1}{2}\int^{+\infty}_{-\infty}dx[(\partial_t\pi(x,t))^2+(\partial_x\pi(x,t))^2+m_\pi^2\pi(x,t)^2].
\label{ham}
\end{equation}

The free particle solutions for $t=0$ are
\begin{eqnarray}
\pi(x,0)=\int^{+\infty}_{-\infty}\frac{dk}{2\pi}\pi_{k0}\,e^{ikx}=
\int^{+\infty}_{-\infty}\frac{dk}{2\pi}\frac{(a_k+a_{-k}^\ast)}{\sqrt{2\omega_k}}e^{ikx}
\nonumber\\
\dot{\pi}(x,0)=\int^{+\infty}_{-\infty}\frac{dk}{2\pi}\xi_{k0}\,e^{ikx}=
\int^{+\infty}_{-\infty}\frac{dk}{2\pi}\sqrt{\frac{\omega_k}{2}}i(a_{-k}^\ast-a_k)e^{ikx}.
\label{sol}
\end{eqnarray}
where $\omega_k=\sqrt{k^2+m_\pi^2}$.

Now, choosing an initial classical density distribution
\cite{Cooper}
\begin{equation}
\rho[\pi,\dot{\pi}]=Z^{-1}(\beta_0)\,
\exp{\left\{-\beta_0\,H[\pi,\dot{\pi}]\right\}}, \label{dens}
\end{equation}
 and substitute the Hamiltonian (\ref{ham})
with the free particle solutions (\ref{sol}), we finally acquire
\begin{equation}
\rho[x_k,y_k]=Z^{-1}(\beta_0)\,
\exp{\left\{-\beta_0\,\int^{+\infty}_{-\infty}\frac{dk}{2\pi}\omega_k(x_k^2+y_k^2)\right\}},
\label{findens}
\end{equation}
with $\beta_0=1/T_0$, and where: $a_k=x_k+iy_k$ with $x_k, y_k$
real. In order to produce a thermal ensemble (at temperature
$T_0$) of configurations for $\pi(x,0)$ and $\dot{\pi}(x,0)$, we
select $x_k$ and $y_k$ from the gaussian distribution
(\ref{findens}), assemble $a_k$ and then substitute in
(\ref{sol}).
 \begin{figure}[!]
\begin{center}
\mbox{\epsfig{figure=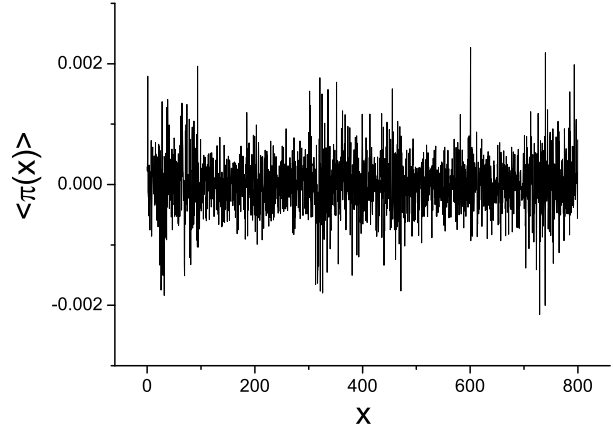,width=9cm,angle=0}}
\caption{\it The $\pi$-field on the lattice averaged over the
ensemble of configurations.} \label{pi0}
 \end{center}
 \end{figure}
 The corresponding $\langle\pi(x)\rangle$ profile is
shown in fig.~\ref{pi0}.
  All the characteristics of the
$\pi$-ensemble such as the correlation function
$\langle\pi(x)\pi(x+\delta
x)\rangle-\langle\pi(x)\rangle\langle\pi(x+\delta x)\rangle$,
which turns out to be a $\delta$-function, are consistent with the
assumption of an ideal thermal gas.

\subsection{Numerical Results. Two coupled fields}

 We solve equations of motion
(\ref{eomsp}) in 1-D 2000-site lattice following the
discretization scheme (\ref{leap-frog}), using for initial
conditions an ensemble of $10^4$ $\s$ and $\pi$ configurations
satisfying the aforementioned requirements, that is fractal $\s$
and conventional $\pi$ (corresponding to dimensionless temperature
$T_0=1$) configurations. As in the single field case we focus on
the evolution of the $\langle\int^{x_0+\xi}_{x_0} |\sigma(x)|
dx\rangle$, which defines a fractal measure, initially having the
characteristic mass dimension $5/6$, generated by the $\s$-field.
\begin{figure}[h]
\begin{center}
\mbox{\epsfig{figure=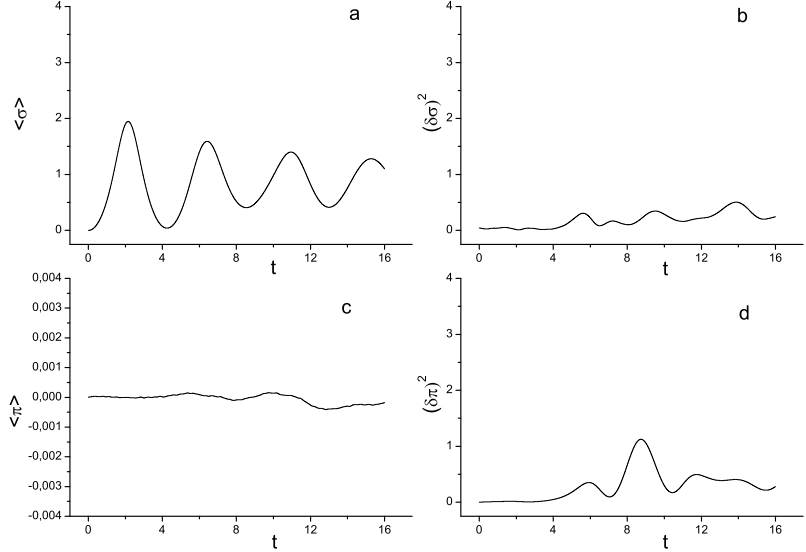,width=12cm,angle=0}}
\caption{\it Mean field and standard deviation evolution in the
two-field case for $A=1$, $\la=1$, in dimensionless units. Mind
the different vertical axis scale in plot c).} \label{spfields}
 \end{center}
 \end{figure}

In figs \ref{spfields}a,b we demonstrate the evolution of the mean
$\s$-field and its standard deviation, and in figures
\ref{spfields}c,d the corresponding quantities for the
$\pi$-field, for $A=1$ and $\la=1$. The absence of a term linear
in $\pi$-field in the potential (\ref{potsp}), leads to slight
oscillations, due to the small initial total energy of the
$\pi$-field, of the mean value $\langle\pi\rangle$ around zero.
The anharmonic character of these oscillations relies on the
non-linear form of the equation of motion (\ref{eomsp}), and is
enhanced relatively to the $\s$ case due to the large variation of
the coupling term $\lambda(\s^2-1)\pi$. Note however that although
the mean field value $\langle\pi\rangle$ remains small, this is
not the case for its fluctuations which can become quite large. On
the other hand, as one can see in figures \ref{spfields}a,b, the
corresponding $\s$ quantities do not differ significantly from the
single field case since $\pi$-field remains small.
\begin{figure}[h]
\begin{center}
\mbox{\epsfig{figure=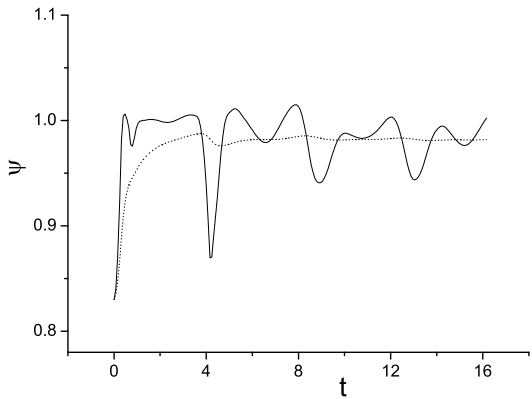,width=8cm,angle=0}}
\caption{\it Time evolution of the slope $\psi$ (solid line) and
its time average $\langle\psi\rangle_t$ (dotted line) in the
two-field case, for $A=1$ and $\la=1$, in dimensionless units.}
\label{sp.power}
 \end{center}
 \end{figure}

In fig.~\ref{sp.power} we depict the slope $\psi(t)$ for the
aforementioned evolution scenario of the $\s$-field interacting
with the $\pi$-field. In general we observe the same phenomenon as
in the single field case, that is $\psi(t)$ becomes approximately
equal to its initial value every time $\langle\s\rangle$
approaches zero. However, the stronger non-linearity forbids
$\langle\s\rangle$ to become exactly zero but rather to obtain a
small finite value, which in turn has as a consequence that
$\psi(t)$ does not approach so closely the initial value as in the
single field case. This is illustrated in figs.~\ref{sp.power} and
~\ref{spfields}a, at $t\sim 4$ and $t\sim9$, which correspond to
the times when $\langle\s\rangle(t)$ has a local minimum. In the
case of the first minimum ($t\sim 4$), $\langle\s\rangle$ returns
closer to zero and therefore $\psi(t)$ becomes almost $5/6$,
compared to the case of the second minimum at $t\sim9$, when
$\langle\s\rangle$ reaches a greater value ($\sim0.5$) and
correspondingly $\psi(t)$ deviates stronger from its initial
value. However, in both cases the system maintains a memory of its
initial fractal characteristics.

We have evolved the system with various $\pi$-field initial
conditions (random, constant etc) in order to check the generality
of the results described above. It turns out that for the
two-field case  the re-establishment of the initial fractal
geometry does not depend on the specific $\pi$-field initial
conditions, unless they are fine tuned so that $\langle\s\rangle$
is driven to large values without occasionally returning close to
the region $\langle\s\rangle\approx0$.

\section{Discussion and Conclusions}

In this work we investigate the evolution of the fractal
characteristics of a single or two coupled scalar fields system,
within the framework of the general $\s$-model. After a relatively
rapid deformation of the initial fractal geometry, we observe that
it is being re-established almost periodically each time the mean
value of the $\s$-field returns to zero. This effect is obtained
using both random, as well as deterministic (Cantor-like) fractal
set up. The key point for the occurrence of this behavior is the
condition of initial equilibrium. As the system practically
consists of coupled anharmonic oscillators, this condition is
expressed through the zero initial kinetic energy for each
oscillator, leading to a synchronous evolution of the entire
system. Additionally, the fractal measure characterizing the
initial state is based on the fact that the corresponding field
configurations represent fluctuations (with a specific pattern)
around zero. Thus, each time the oscillators pass through their
turning points, associated to zero mean field value, in phase, the
initial fractal geometry is being re-established. The above
evolution, at least qualitatively, is quite general and robust for
a very wide parametric space.

The scenario analyzed in this paper could be extended to 3
dimensions and serve as a model in order to describe the evolution
of the order parameter fluctuations in a critical system. In
particular it could be used to explore the non-conventional
correlations which are expected to occur during the formation of
an isoscalar condensate in a heavy ion collision experiment. In
this case one has to adapt the $\s$-model Lagrangian in order to
describe correctly the characteristics of the order parameter
associated with the 2nd order critical end point of the chiral QCD
phase transition \cite{RW}. At the phenomenological level these
correlations are expressed through the fractal mass dimension of
the $\s$-field configurations, determining the statistical
properties of the condensate. The dynamics of the system is
similar to the two-field case investigated above, provided that
the critical system is initially at equilibrium. One can study the
evolution of the initial fractal characteristics of the $\s$-field
and the possibility to leave signals at the detectors, supplying
an indication of the phase transition. The discussion of the
present work supports this eventuality since the time average (in
order to simulate the experimental conditions) of the periodical
deformation and re-establishment of the initial fractal geometry,
will leave a signature. The specific application of the present
work in the case of QCD phase transition has been performed in \cite{Futurework}.\\

\paragraph*{{\bf{Acknowledgements:}}} We thank V. Constantoudis and
N. Tetradis for useful discussions. One of us (E.N.S) wishes to
thank the Greek State Scholarship's Foundation (IKY) for financial
support. The authors acknowledge partial financial support through
the research programs ``Pythagoras'' of the EPEAEK II (European
Union and the Greek Ministry of Education) and ``Kapodistrias'' of
the University of Athens.

\end{document}